\documentstyle[aps,multicol,prl,epsf]{revtex}
\begin{document}
\draft

\title{Measuring the transmission phase of a quantum dot in a closed interferometer}

\author{%
Amnon Aharony$^a$, Ora Entin-Wohlman$^{a,b}$,
and Yoseph Imry$^c$}

\address{%
$^a$ School of Physics and Astronomy, Raymond and Beverly Sackler
faculty of Exact Sciences, \\Tel Aviv University, Tel Aviv 69978,
Israel\\
$^b$ Albert Einstein Minerva Center for Theoretical Physics,
Weizmann Institute of Science, Rehovot 76100, Israel\\
$^c$ Department of Condensed Matter Physics, Weizmann Institute of
Science, Rehovot 76100, Israel.}

\date{\today}
\maketitle

\begin{abstract}
 The electron transmission through a {\it closed}
Aharonov-Bohm mesoscopic solid-state interferometer, with a
quantum dot (QD) on one of the paths, is calculated exactly for a
simple model. Although the conductance is an even function of the
magnetic flux (due to Onsager's relations), in many cases one can
use the measured conductance to extract both the amplitude and the
phase of the ``intrinsic" transmission amplitude
$t_D=-i|t_D|e^{i\alpha_D}$ through the ``bare" QD. We also propose
to compare this indirect measurement with the (hitherto untested)
direct relation $\sin^2(\alpha_D) \equiv |t_D|^2/\max(|t_D|^2)$.

\end{abstract}


\begin{multicols}{2}


Recent advances in nanoscience raised interest in quantum dots
(QDs), which represent artificial atoms with experimentally
controllable properties\cite{review,book}.  Connecting the QD via
metallic leads to electron reservoirs yields resonant transmission
through the QD, with peaks whenever the Fermi energy in the leads
crosses a resonance on the QD. The energies of the latter are
varied by controlling the plunger gate voltage on the QD, $V$. The
quantum information on the tunneling of an electron is contained
in the complex transmission amplitude, $t_{D}=-i\sqrt{{\cal
T}_{D}}e^{i\alpha_D}$. The phase $\alpha_D$ is particularly
interesting, given its relation to the additional electron
occupation in the system via the Friedel sum rule
\cite{friedel,langreth}. This phase is also predicted to exhibit
interesting behavior e.g. near a Kondo-like
resonance\cite{hewson}. This motivated experimental attempts to
measure $\alpha_D$ \cite{yacoby,schuster}, using the Aharonov-Bohm
interferometer (ABI)\cite{azbel}.

In the ABI, the QD is placed on one branch, in parallel to a
``reference" branch (both connecting the two external leads).  A
magnetic flux $\Phi$ in the area between the two branches creates
a phase difference $\phi=e\Phi/\hbar c$ between the wave functions
in the two branches\cite{AB}.  In the {\it two-slit limit}, the
total ABI transmission is
\begin{eqnarray}
{\cal T}=|t|^2=|t_De^{i\phi}+t_B|^2=A+B\cos(\phi+\beta),
\label{2slit} \end{eqnarray} with $\beta=\alpha_D-\kappa$, where
$\kappa$ contains $V$-independent contributions from the reference
transmission, $t_B=-i|t_B|e^{i\delta_B}$, and from the electron
``optical" paths on the two branches. However, for the ``closed"
two-terminal geometry, unitarity (conservation of current) and
time reversal symmetry imply the Onsager relations \cite{onsager},
which state that the two-terminal conductance, ${\bf
G}=(e^2/h){\cal T}$, is an {\it even function} of $\phi$.
Therefore, a naive fit of the experimental transmission to Eq.
(\ref{2slit}) {\it must} yield $\beta=0$ or $\pi$ -- with no
relation to $\alpha_D$. Indeed, the experimental data\cite{yacoby}
for ${\cal T}$ depend only on $\cos\phi$\cite{jpn}.

Aiming to measure a {\it non-trivial} AB phase shift $\beta$ then
led to experiments with ``open" interferometers
\cite{schuster,ji}, which contain additional ``leaky" channels,
breaking the Onsager symmetry. A fit to Eq. (\ref{2slit}) then
yields a phase $\beta$ which increases (with $V$) {\it gradually}
from 0 to $\pi$ through each resonance. However, the detailed
$V$-dependence of $\beta$ {\it depends on the strength} of the
coupling to the additional terminals \cite{prl}. Although it is
possible to optimize this strength, and reproduce the two-slit
conditions \cite{bih}, this involves large uncertainties.

In the present paper we present exact results for the total
transmission of the closed ABI, ${\cal T}$. Although ${\cal T}$ is
even in $\phi$, contradicting the simple two-slit Eq.
(\ref{2slit}), it {\it does} depend on {\it both} ${\cal T}_D$ and
$\alpha_D$.  Under appropriate conditions (see below), one can
thus extract $\alpha_D$ from the measured ${\cal T}$, {\it
eliminating the need to open the interferometer}. This possible
extraction was not noticed in earlier discussions of the closed
ABI. Theoretical analyses used the Keldysh technique, combined
with the wide-band and related approximations\cite{bulka,H}, or
ignored electron-electron interactions\cite{hartzstein}. These
approximations, which sometimes miss important features of the
results (see below), are avoided in our calculation, which is done
in the linear response limit, and at temperature $T=0$.

\begin{figure}[h]
\hspace{.8cm}\leftline{\epsfclipon\epsfxsize=2.5in\epsfysize=1.3in\epsffile{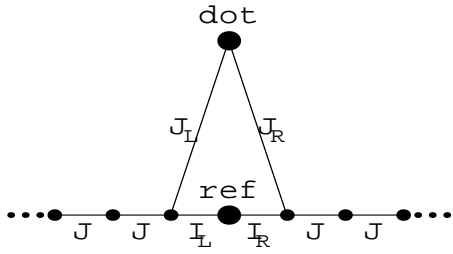}}
\vspace{.5cm} \caption{Model for the closed ABI.}
\label{fig1}
\end{figure}

 We demonstrate our results for a simple
lattice model, shown in Fig. \ref{fig1}: for $\Phi=0$, each (unit
length) segment in the figure represents a real tight-binding
hopping matrix element $-J,~-I_L,~-I_R,~-J_L$ and $-J_R$, as
indicated. All the on-site energies are zero, except $\epsilon_D$
on the site ``dot" and $\epsilon_0$ on the site ``ref" (which sits
on the reference path and represents a simple point contact, a
tunnel junction, etc). The latter two energies can be varied
experimentally by the plunger (or point contact) gate voltages
$\epsilon_D \equiv V$ and $\epsilon_0 \equiv V_0$ \cite{jpn}. As
usual for such models, electron-electron interactions are included
only via an on-site Hubbard interaction $U$ on the QD. The AB
phase in the triangle, $\phi =\phi_L+\phi_R$, is included by
attaching a factor $e^{i\phi_L}$ ($e^{i\phi_R}$)  to the hopping
matrix element $J_L$ ($J_R$). At $T=0$, the electron energy
$\epsilon_k=-2J \cos k$ is equal to the Fermi energy on the leads,
$\epsilon_F$, and we calculate the transmission for electrons with
spin $\sigma$.

We start by reviewing the ``intrinsic" transmission through the
QD, without the reference path (e.g. for large
$|V_0|=|\epsilon_0|$, or with $I_L=I_R=0$). Adapting the results
of Ref. \cite{ng}, one has
\begin{eqnarray}
t_D=-i \gamma_D\sin \alpha_D e^{i\alpha_D} \equiv 2i \sin |k|
J_LJ_R g_{D}(\epsilon_k)/J, \label{td}
\end{eqnarray}
with the QD asymmetry factor $\gamma_D=2J_LJ_R/(J_L^2+J_R^2)$ and
the ``intrinsic" Green function on the QD,
$g_{D}(\epsilon_k)=1/[\epsilon_k-\epsilon_D-\Sigma_D(\epsilon_k)]$.
Here, $\Sigma_D(\epsilon_k)$ is the self-energy on the QD, which
contains contributions from the leads,
$\Sigma_{D,ext}=-e^{i|k|}(J_L^2+J_R^2)/J$ (which exists also for
the non-interacting case \cite{prl}), and from the
electron-electron interactions on the QD itself,
$\Sigma_{D,int}(\omega)$ (which vanishes when $U=0$).
As $\epsilon_D \equiv V$ increases, $\alpha_D$ grows gradually
from zero (far below the resonance), through $\pi/2$ (at the
resonance), towards $\pi$ (far above the resonance).

Interestingly, for this one-dimensional model, normalizing the
measured ${\cal T}_D=|t_D|^2=\gamma_D^2\sin^2(\alpha_D)$ by its
($V$-independent) maximum $\gamma_D^2$ yields the value of
$\alpha_D$. Assuming coherence, this (hitherto ignored) method for
measuring $\alpha_D$ {\it directly} from ${\cal T}_D$ {\it
eliminates the need for any complicated interferometry}!
\cite{com1} In the remainder of this paper we discuss ways of
extracting $\alpha_D$ {\it indirectly}, from the {\it closed} ABI
measurements. Comparing results from $\sin^2(\alpha_D)={\cal
T}_D/\gamma_D^2$, from the closed ABI (below) and from the open
ABI \cite{bih} (all with the same QD) should serve as {\em
consistency checks} for this conclusion.

The same analysis yields the transmission amplitude through the
reference path (when e.g. $J_L=J_R=0$),
\begin{eqnarray}
t_B=-i \gamma_B\sin \delta_B e^{i \delta_B} \equiv 2i \sin
|k|I_LI_Rg_B(\epsilon_k)/J \label{tb}
\end{eqnarray}
with the bare reference site Green function
$g_B=1/[\epsilon_k-\epsilon_0+e^{i|k|}(I_L^2+I_R^2)/J]$, and
$\gamma_B=2I_LI_R/(I_L^2+I_R^2)$.
In the two-slit situation, Eqs. (\ref{td}) and (\ref{tb}) suffice
to determine the overall transmission, as in Eq. (\ref{2slit}).
However, the situation is more complicated for the closed ABI. The
main result of the present paper concerns the {\it exact}
transmission amplitude through the closed ABI,
\begin{eqnarray}
t=A_Dt_D e^{i \phi}+A_B t_B, \label{tt1}
\end{eqnarray}
where we find $A_D=
g_B(\epsilon_k-\epsilon_0)G_D(\epsilon_k)/g_D(\epsilon_k)$ and
 $A_B=1+G_D(\epsilon_k)\Sigma_{ext}(\epsilon_k)$.
Here, $G_{D}(\omega )=1/[\omega -\epsilon_{D}-\Sigma(\omega )]$ is
the fully ``dressed" Green function on the QD, with  the dressed
self-energy $\Sigma=\Sigma_{int}+\Sigma_{ext}$. Both terms here
differ from their counterparts in the ``intrinsic" $\Sigma_D$, by
contributions due to the reference path. Equation (\ref{tt1})
looks like the two-slit formula, $t=t_D e^{i\phi}+t_B$. However,
each of the terms is now {\em renormalized}: $A_D$ contains all
the additional processes in which the electron ``visits" the
reference site ($A_D=1$ when $I_L=I_R=0$), and $A_B$ contains the
corrections to $t_B$ due to ``visits" on the dot. In fact, a
physical derivation of Eq. (\ref{tt1}) amounts to starting from
Eq. (\ref{td}), and adding an infinite power series in $I_L$ and
$I_R$. We now discuss the $\phi$-dependence of  ${\cal T} \equiv
|t|^2$, in connection with the Onsager relations and with the
possible indirect extraction of $\alpha_D$.

We first note that
both parts in $\Sigma(\epsilon_k)$ are even in $\phi$, due to
additive contributions (with equal amplitudes) from clockwise and
counterclockwise motions of the electron around the ring (see e.g.
Refs. \onlinecite{azbel,prl,hartzstein,W}). In order that ${\cal
T}$ also depends only on $\cos\phi$, as required by the Onsager
relations, the ratio $K\equiv A_Bt_B/(A_Dt_D) \equiv \tilde
x[G_{D}(\epsilon_k)^{-1}+\Sigma_{ext}(\epsilon_k)]$, with the real
coefficient $\tilde x=I_LI_R/[J_LJ_R(\epsilon_k-\epsilon_0)]$,
must be real, i.e.
\begin{eqnarray}
&\Im[G_{D}(\epsilon_k)^{-1}+\Sigma_{ext}(\epsilon_k)] \equiv \Im
\Sigma_{int} \equiv 0. \label{real}
\end{eqnarray}
The same relation follows from the unitarity of the $2 \times 2$
scattering matrix of the ring. This relation already appeared for
the special case of single impurity scattering, in connection with
the Friedel sum rule\cite{langreth}, and was implicitly contained
in Eq. (\ref{td}), where $\Im \Sigma_{D,int}=0$ \cite{ng}.
Equation (\ref{real}) implies that (at $T=0$ and
$\omega=\epsilon_k$) {\it the interaction self-energy}
$\Sigma_{int }(\epsilon_k)$ is {\it real}, and therefore the width
of the resonance, $\Im G_{D}(\epsilon_k)^{-1}$,
is {\it fully determined by the non-interacting self-energy} $\Im
\Sigma_{ext}(\epsilon_k)$.

Since $\Sigma_{ext}(\omega)$ depends only on the (non-interacting)
tight-binding terms, it is easy to calculate it explicitly. We
find
$\Sigma_{ext}(\epsilon_k)=\Sigma_{D,ext}(\epsilon_k)+\Delta_{ext}$,
where
\begin{eqnarray}
\Delta
_{ext}=e^{2i|k|}g_B(J_L^2I_L^2+J_R^2I_R^2+2J_LJ_RI_LI_R\cos\phi)/J^2.
\label{A}
\end{eqnarray}
The  term proportional to $\cos\phi$ comes from the electron
clock- and counterclockwise motion around the ABI ``ring".
Similarly, one can write $\Sigma_{int}(\epsilon_k)=\Sigma_{
D,int}(\epsilon_k)+\Delta_{int}$, and thus
$G_{D}(\epsilon_k)^{-1}=g_{D}(\epsilon_k)^{-1}-\Delta$, with
$\Delta=\Delta_{ext}+\Delta_{int}$.  Hence,
$t=A_Dt_D(e^{i\phi}+K)$. Writing also
$A_D=C/[1-g_D(\epsilon_k)\Delta]$, with
$C=(\epsilon_k-\epsilon_0)g_B$, we find
\begin{eqnarray}
{\cal T}=|C|^2{\cal T}_D \frac{1+K^2+2 K\cos\phi}{1-2\Re[g_D
\Delta]+|g_D\Delta|^2}. \label{TT2}
\end{eqnarray}


 Equation (\ref{TT2}) presents an alternative form of our main
 result. Although the numerator looks like the two-slit Eq.
 (\ref{2slit}), with $\beta=0$ or $\pi$ (depending on ${\rm sign} K$),
 the new physics is contained in the denominator -- which
 becomes important in the vicinity of a resonance.
The central term in this denominator depends explicitly on  the
phase of the complex number $g_D$. Since this number is directly
related to $t_D$, via Eq. (\ref{td}), one may expect to extract
$\alpha_D$ from a fit to Eq. (\ref{TT2}), taking advantage of the
dependence of the denominator on $\cos\phi$.
 Physically, this dependence
originates from the infinite sum over electron paths which
circulate the ABI ring.
 The rest of this paper is devoted to the conditions for such an extraction.
  Generally, this is not trivial, as one needs
the detailed dependence of $\Delta$ on $\cos\phi$ and on the
various parameters. We have presented this dependence for
$\Delta_{ext}$, but not for $\Delta_{int}$.

The extraction of $\alpha_D$ becomes easy when one may neglect
$\Delta_{int}$. The simplest case for this is for single-electron
scattering, when $\Sigma_{int}=0$. Interactions (i.e. $U$) are
also negligible for a relatively {\it open} dot, with small
barriers at its contacts with the leads\cite{mat}.
Another effectively single-electron scattering case arises near a
Coulomb blockade resonance, when the effect of interactions can
simply be absorbed into a Hartree-like shift,
$\epsilon_D+\Sigma_{int} \rightarrow \epsilon_D+N U$, if one {\it
assumes} that $N$ depends smoothly on the number of electrons on
the QD, and not on $\phi$\cite{W}. If one may neglect
$\Delta_{int}$, then $\Delta \approx \Delta_{ext}$ is given in Eq.
(\ref{A}). Using also Eqs. (\ref{td}) and (\ref{tb}), we find
\begin{eqnarray}
{\cal T}=|C|^2{\cal T}_D\frac{1+K^2+2
K\cos\phi}{1+2P(z+\cos\phi)+Q(z+\cos\phi)^2}, \label{ttt}
\end{eqnarray}
where $z=(J_L^2I_L^2+J_R^2I_R^2)/(2J_LJ_RI_LI_R)$, $P=\Re
[vt_Bt_D]$, $Q=|vt_B|^2{\cal T}_D$, and $v=e^{2i|k|}/(2\sin^2|k|)$
depends only on the Fermi wavevector $k$, independent of any
detail of the ABI. A 5-parameter fit to the explicit
$\phi$-dependence in Eq. (\ref{ttt}) for given values of $V$ and
$V_0$ then yields $|C|^2{\cal T}_D,~K,~z,~P$ and $Q$, and thus
$\cos(\alpha_D+\delta_B+2|k|)=P/\sqrt{Q}$, from which one can
extract the $V$-dependence of $\alpha_D$. The same $V$-dependence
of $\alpha_D$ is also contained in $K \propto
(\cot\alpha_D+\cot|k|)$). As discussed after Eq. (\ref{td}), our
model also implies that ${\cal T}_D=\gamma_D^2\sin^2(\alpha_D)$.
Since the $V$-dependence of ${\cal T}_D$ can also be extracted
from the fitted values of either $|C|^2{\cal T}_D$ or $Q$, we end
up with several consistency checks for the determination of
$\alpha_D$. Additional checks arise from direct measurements of
${\cal T}_D$ and ${\cal T}_B=|t_B|^2$, by taking the limits
$|V_0|=|\epsilon_0| \rightarrow \infty$ or $|V|=|\epsilon_D|
\rightarrow \infty$.

\begin{figure}[h]
\leftline{\epsfclipon\epsfxsize=3.6in\epsfysize=2.5in\epsffile{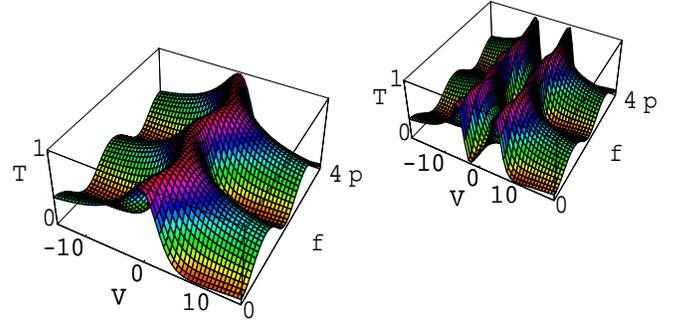}}
\vspace{.3cm} \caption{AB transmission ${\cal T}$ versus the AB
phase $\phi$ and the gate voltage $V$, for one (LHS) and two (RHS)
non-interacting resonances.}\label{fig2}
\end{figure}

The LHS frame in Fig. \ref{fig2} shows an example of the $V$- and
$\phi$-dependence of ${\cal T}$ for this limit (no interactions),
with $k=\pi/2$ and $J_L=J_R=I_L=I_R=1,~V_0=4$ (in units of $J$),
implying
 $K=\epsilon_D/\epsilon_0=V/V_0$.  Far away from the resonance
${\cal T} \ll 1$, $Q \ll |P| \ll 1$ and $|K| \gg 1$, yielding the
two-slit-like form  ${\cal T} \approx A+B\cos\phi$, dominated by
its first harmonic, with $B/A \approx 2[K^{-1}-P]$. However, close
the the resonance ${\cal T}$ shows a rich structure; the
denominator in Eq. (\ref{ttt}) generates higher harmonics, and the
two-slit formula is completely wrong. This rich structure may be
missed if one neglects parts of the $\phi$-dependence of $\Delta$,
as done in Ref. \cite{H}. Note also the Fano vanishing \cite{jlt}
of ${\cal T}$ for $V \sim 10$ at $\phi=2n\pi$, with integer $n$.
Without interactions, everything can be extended to a QD with many
resonances, e.g. due to Coulomb blockade shifts in the effective
$\epsilon_d$ with the number of electrons. Using a generalization
to Eq. (\ref{ttt}), given in Ref. \cite{bih}, the RHS frame in
Fig. \ref{fig2} shows results for two resonances, with
$\epsilon_D=\pm 5$. Interestingly, Fig. \ref{fig2} is
qualitatively similar to the experimentally measured transmission
in Ref. \onlinecite{jpn}. However, so far there has been no
quantitative analysis of the experimental data.

To treat the general case, we need information on $\Delta_{int}$.
First of all, we emphasize that {\it a successful fit to Eq.
(\ref{ttt}) justifies the neglect of the} $\phi$-{\it dependence}
of $\Delta_{int}$. If the various procedures to determine
$\alpha_D$ from Eq. (\ref{ttt}) yield the same $V$-dependence,
this would also confirm that $\Delta_{int}$ is negligibly small. A
failure of this check, or a more complicated dependence of the
measured ${\cal T}$ on $\cos\phi$, would imply that $\Delta_{int}$
is not negligible.

As seen from Eq. (\ref{A}), $\Delta_{ext}$ is fully determined by
a single ``visit" of the electron at ``ref". For small ${\cal
T}_B$, or large $|V_0|=|\epsilon_0|$, it is reasonable to
conjecture that $\Delta_{int}$ is also dominated by such
processes. In that case, we expect $\Delta_{int}$ to be
proportional to the same brackets as in Eq. (\ref{A}), i.e.
$\Delta_{int} \approx w(z+\cos\phi)$, with a real coefficient $w$.
This yields the same dependence of ${\cal T}$ on $\cos\phi$ as in
Eq. (\ref{ttt}), with a shifted coefficient $v$. If $w$ depends
only weakly on $V$, then this shift has little effect on the
determination of $\alpha_D$. Again, the validity of this approach
relies on getting the same $V$-dependence of $\alpha_D$ from all
of its different determinations.

The situation becomes more complicated near a Kondo-like
resonance. Maintaining the (non-trivial) assumption that
$G_{D}=1/[\omega-\epsilon_D-\Sigma_D(\omega)]$, the Kondo peak at
the Fermi energy must be generated by $\Sigma_{D}$. For the
intrinsic QD, this yields $\alpha_D=\pi/2$ and $t_D=\gamma_D$,
resulting in a $V$-independent plateau for ${\cal T}_D$. {\it A
priori}, it is not obvious what happens in the presence of the
``reference" path. Hofstetter {\it et al.} identified the Kondo
region by requiring that the phase $\delta_{\rm res}$ of the fully
dressed Green function $G_D$ be equal to $\pi/2$. Our result for
$G_D$ shows that this is impossible: the phase $\delta_{\rm res}$
{\it depends} on $\phi$, via the $\phi$-dependence of $\Delta$,
and thus {\it cannot} be set at the constant value $\pi/2$.
(Apparently, this $\phi$-dependence was neglected in the analytic
parts, and weak for the numerical parameters used in Ref.
\cite{H}). Alternatively, one might assume that the ``bare" QD
sticks to the Kondo resonance, and thus $\alpha_D=\pi/2$
(independent of $V$) even in the ABI. Equation (\ref{TT2}) then
replaces the Kondo plateau by a complicated dependence on $\phi$
(including the first harmonic), which differs significantly from
that of Ref. \cite{H}. Clearly, this limit requires more research.

Finally, we give some more details of our derivation. Our
Hamiltonian, which simply adds the reference path to that of Ng
and Lee \cite{ng}, is
\begin{eqnarray}
{\cal
H}&=&\epsilon_{D}\sum_{\sigma}d^{\dagger}_{\sigma}d_{\sigma}+
\frac{U}{2}\sum_{\sigma}n_{d\sigma}n_{d\overline{\sigma}}\nonumber\\
&+&\sum_{k\sigma}\epsilon_{k}c^{\dagger}_{k\sigma}c_{k\sigma}
+\sum_{k\sigma}\Bigl ({\cal
V}_{k}d^{\dagger}_{\sigma}c_{k\sigma}+{\cal
V}^{\ast}_{k}c^{\dagger}_{k\sigma}d_{\sigma}\Bigr )\nonumber\\
&+&\epsilon_{0}\sum_{\sigma}c^{\dagger}_{0\sigma}c_{0\sigma}
+\sum_{k\sigma}\Bigl ({\cal U}_k
c^{\dagger}_{0\sigma}c_{k\sigma}+{\cal U}_k^\ast
c^{\dagger}_{k\sigma}c_{0\sigma}\Bigr ), \label{HH}
\end{eqnarray}
where $c_{k\sigma}^\dagger$ creates single particle eigenstates
(with spin $\sigma$) on the unperturbed ``background" chain (with
$I_L=I_R=J$, $J_L=J_R=0$), with eigenenergy $\epsilon_{k}=-2J\cos
k$, while
$c_{0\sigma}=\sum_{k}c_{k\sigma}/\sqrt{N}$, ${\cal
U}_k=-[(I_L-J)e^{-ik}+(I_R-J)e^{ik}]/\sqrt{N}$,
 and
${\cal V}_{k}=-(J_Le^{i\phi_{\ell}-ik}+J_{R}e^{-i\phi_{r}+ik}
)/\sqrt{N}$.
The operators on the dot, $d_{\sigma}$ and $d^{
\dagger}_{\sigma}$, anti-commute with
$c_{k\sigma},c^{\dagger}_{k\sigma}$. Also,
$n_{d\sigma}=d^{\dagger}_{\sigma}d_{\sigma}$,
and $\overline{\sigma} \equiv -\sigma$.

  As stated above, one can derive Eq. (\ref{tt1}) by expanding Eq. (\ref{td})
  in powers of $I_L$ and $I_R$. A more general approach uses the
  standard relation between the
 $2 \times 2$ scattering matrix $T_{kk'}$
and the matrix of retarded single-particle Green functions,
$G_{kk'}(\omega)=\delta_{kk'}g_{k}^0+g^0_kT^\sigma_{kk'}g^0_{k'}$,
with $g^0_{k}(\omega)=1/(\omega-\epsilon_k)$, evaluated on the
energy shell,
$\omega=\epsilon_F=\epsilon_k=\epsilon_{k'}$\cite{hewson}. The
equation-of-motion (EOM) method is then used to express
$(\omega-\epsilon_k)G_{kk'}(\omega)$ and
$(\omega-\epsilon_k)G_{kd}(\omega)$ as linear combinations of each
other and of $G_D(\omega)$, allowing us to express each of them
(and thus also $t \propto T_{|k|,|k|}$) in terms of
$G_{D}(\omega )$, yielding Eq. (\ref{tt1}). Since we do not use an
explicit solution for $G_{D}(\omega )$ itself, we don't need to
deal with the higher order correlation functions (due to $U$),
which appear in its EOM.

We hope that our paper will stimulate attempts to fit experimental
data to our Eq. (\ref{ttt}), and to compare the resulting
$\alpha_D$ with its direct estimate via ${\cal T}$. This procedure
should work in many cases. We also hope that our paper will
stimulate more detailed theoretical calculations of
$\Delta_{int}$. As explained, the existing approximate
calculations miss the crucial $\phi$-dependence of these
interaction-dependent terms.

We thank R. Englman, M. Heiblum, Y. Levinson, A. Schiller, H. A.
Weidenm\"uller and A. Yacoby for helpful conversations. This
project was carried out in a center of excellence supported by the
Israel Science Foundation, with additional support from the
German-Israeli Foundation (GIF), and from the German Federal
Ministry of Education and Research (BMBF) within the Framework of
the German-Israeli Project Cooperation (DIP).

\newpage

\end{multicols}
\end{document}